# Uniform heating of materials into the warm dense matter regime with laser-driven quasi-monoenergetic ion beams

W. Bang,[1, a)] B. J. Albright,[1] P. A. Bradley,[1] E. L. Vold,[1] J. C. Boettger,[1] and J. C. Fernández[1]

[1]*Los Alamos National Laboratory, Los Alamos, New Mexico 87545, USA*

**Abstract**

In a recent experiment on the Trident laser facility, a laser-driven beam of quasi-monoenergetic aluminum ions was used to heat solid gold and diamond foils isochorically to 5.5 eV and 1.7 eV, respectively. Here theoretical calculations are presented that suggest the gold and diamond were heated uniformly by these laser-driven ion beams. According to calculations and SESAME equation-of-state tables, laser-driven aluminum ion beams achievable on Trident, with a finite energy spread of $\Delta E/E \sim 20\%$, are expected to heat the targets more uniformly than a beam of 140 MeV aluminum ions with zero energy spread. The robustness of the expected heating uniformity relative to the changes in the incident ion energy spectra is evaluated, and expected plasma temperatures of various target materials achievable with the current experimental platform are presented.

---

Author to whom correspondence should be addressed. Electronic mail:
a) wbang@lanl.gov

## I. Introduction

Rapid heating of matter using a short (ps to tens of fs) laser pulse is an emerging research area in plasma physics [1-18]. In these settings, the interaction between intense light and matter is sufficiently strong and fast for the atoms to be ionized instantaneously and the resulting electrons or ions (or both) attain large kinetic energies very quickly. By directly illuminating a small, reduced-mass solid target with an intense laser beam, target temperatures on the order of 100 eV has been demonstrated [16]. Using even smaller targets (~10 nm radius spheres) [19], an ion temperature exceeding tens of keV has been achieved in the laboratory [10], which is sufficiently high for deuterium ions to produce nuclear fusion reactions efficiently [20-23].

While such high temperature plasmas have been reliably produced using this direct and intuitive approach for rapid heating of a target, this approach is less desirable when heating uniformity is crucial. In direct laser heating, most of the laser pulse energy is absorbed by the electrons at the front surface of the target (within 100 nm), and the rest of the target is subsequently heated by the resulting hot electrons and the return currents of cold electrons [24]. A recent study by Lévy *et al*. [14] showed that uniform heating can be achieved for a 0.5 μm thick silver target using a free electron x-ray laser beam, but a uniform heating of a thicker target (>1 μm) has yet to be reported using this approach.

With the development of new laser-driven ion sources in recent years [25-36], an alternative approach was explored. When a beam of energetic (> 1 MeV) ions is incident on a cold target, the ions can transfer a significant amount of their kinetic energy to the target. This heating occurs sufficiently rapidly (~20 ps) [37] that the target does not have time to expand hydrodynamically during heating, and is thus called isochoric heating. In isochoric heating experiments with laser-driven ions [4-8,37-41], the temperature of the target may reach 1–100 eV while still maintaining near-solid density, and the originally cold target becomes warm dense matter (WDM) [42]. For example, Patel *et al.* demonstrated that a laser-driven proton beam can volumetrically heat a 10 μm thick Al target to 23 eV [4], and a subsequent experiment delivering more laser energy showed a rear-surface temperature as high as 83 eV [38].

Although WDM is commonly found in astrophysics (e.g., in planetary cores) as well as in high energy density physics experiments [17,43-47], the properties of WDM are not well understood and are difficult to predict theoretically [8,48,49]. This is because neither the approximations made to describe condensed matter, nor those made to describe high-temperature plasmas are strictly valid in this intermediate regime. WDM samples that are uniformly heated would be ideal for these studies, but have been unavailable to date. Typical laser-driven ion beams possess an exponential energy spectrum, which leads to a lack of uniform heating across the whole target depth because of the presence of a large number of low-energy ions, which tend to heat the front surface of the target more strongly [6,7,30,39,41].

Recently [37], we proposed that this issue could be resolved by the use of ion beams with a quasi-monoenergetic energy spectrum [29,50-54]. In this paper, we examine the details of the expected heating calculations from laser-driven quasi-monoenergetic aluminum ion beams. We study the expected heating uniformity within the samples using a Monte Carlo simulation code, SRIM [55,56], and using SESAME [57] equation-of-state (EOS) tables. We also investigate how sensitive the uniform heating conditions are to the energy spectrum of the incident aluminum ions by using different ion energy spectra in the simulation. Using stopping power calculations and available SESAME EOS tables for various target materials, we have calculated expected plasma temperatures that can be achieved with our experimental platform.

## II. Stopping power calculations

The experiments in Ref. [37] were performed on the Trident laser facility at Los Alamos National Laboratory (LANL), which delivered 60–80 J, 650 fs, 1054 nm wavelength pulses to irradiate 110 nm thick aluminum foils. Figure 1 shows the schematic of the target layout inside a vacuum target chamber. Using an f/3 off-axis-parabola, the peak intensity of the laser pulse on the thin aluminum foil (ion source, not shown in Fig. 1) was about $2\times10^{20}$ W/cm$^2$. The laser-driven aluminum ion beam diverged with a 20° cone half-angle [54], and impinged upon gold and diamond foils located 2.37 mm from the ion source inside the vacuum target chamber. A 5 μm thick aluminum filter, inserted 0.37 mm behind the source and 2.0 mm before the target, blocked any laser light propagating through the 110 nm aluminum foil after it became relativistically transparent [54], ensuring the target was indeed heated isochorically with the laser-driven aluminum ion beam. The filter also blocked low energy protons (< 0.5 MeV) and aluminum ions (<10 MeV). We estimate the heating from laser-generated protons [54] to be insignificant because the stopping powers of gold and diamond for protons are much smaller than those for aluminum ions. For example, the stopping powers of gold and diamond for a 5 MeV proton are less than 1% of those for a 140 MeV aluminum ion.

Figure 1 shows a 10 μm thick gold foil (right hand side) and a 15 μm thick diamond foil (on the left side) with a vacuum gap (~100 μm) between. The aluminum ions (depicted as black arrows in the figure) are incident on the target at 45°. At a source-to-target distance of 2.37 mm, the ions incident on target are nearly parallel to one other. The aluminum ions that pass through the vacuum gap are recorded by a magnetic ion spectrometer [58], which monitors shot-to-shot fluctuations in the incident ion energy spectra and fluence. A Thomson parabola ion spectrometer replaced the magnetic ion spectrometer on some shots.

Figure 2 shows the calculated energy spectra of the transmitted aluminum ions using the SRIM code. The black bars indicate the input data to SRIM, which represent the incident energy spectrum of 10,000 aluminum ions on gold or diamond. The average kinetic energy of the incident ions is 140 (±33) MeV. This input energy spectrum is based on a typical energy spectrum of the aluminum ions [54] measured from a Thomson parabola ion spectrometer. The red bars indicate the calculated energy spectrum of the aluminum ions after penetrating through a 10 μm thick gold foil at 45°. Likewise, the blue bars indicate the calculated energy spectrum of the aluminum ions after penetrating through a 15 μm thick diamond foil at 45°. The aluminum ions that went through the vacuum gap did not lose their kinetic energy, and were measured experimentally in Ref. [37], showing close agreement with the input to SRIM (black bars). According to these simulations, 82% of the incident aluminum ions are expected to penetrate the 10 μm thick gold foil, while 86% of the incident aluminum ions are expected to exit the 15 μm thick diamond foil. The average kinetic energy of aluminum ions after traversing a 10 μm thick gold foil is expected to be 42 (±29) MeV, while the average kinetic energy of the ions after traversing a 15 μm thick diamond foil is expected to be 68 (±35) MeV. Indeed, a large fraction of their original ion beam kinetic energy is expected to be transferred to the gold and diamond foils according to our SRIM simulations.

Although the SRIM code uses cold stopping powers of gold and diamond, we expect the uncertainties in these calculations would be reasonably small. In a recent paper [47], Zylstra *et al*. reported their experimental results showing that the stopping power of warm dense Be at 32 eV for 14.7 MeV protons was merely 3–8% higher than the cold stopping power predicted from SRIM. Since the stopping power is known to increase very gradually with the ionization level of the target [47] and the calculated temperatures of our warm dense plasmas are only 5.5 eV for gold and 1.7 eV for diamond, we expect minor increases of the ionization level and estimate the errors in our SRIM calculations would be smaller than 10%.

In Fig. 3(a), the stopping power of gold is plotted as a function of target depth. A flat curve in this plot means a very uniform heating across the whole target depth. Solid red circles indicate the average stopping power of gold for the laser-generated aluminum ions, while hollow black triangles show the stopping power of gold for a perfectly monoenergetic 140 MeV aluminum ion. These SRIM simulations suggest that the laser-driven quasi-monoenergetic aluminum ion beam with a finite energy spread heats the 10 μm thick gold foil even more uniformly than a perfectly monoenergetic aluminum ion beam, a result of a fortuitous balance between heating power from ions in the low-energy part of the spectra absorbed in the target (decreasing stopping power with target depth) and heating from ions from the high-energy part (increasing stopping power with target depth). A target depth of 14 µm in Fig. 3(a) is equivalent to the travel distance of ions within a 10 µm thick gold at 45°.

Figure 3(b) shows the stopping power of diamond as a function of target depth. Solid blue triangles represent the average stopping power of diamond for the laser-generated aluminum ions, while hollow black triangles indicate the stopping power of diamond for a perfectly monoenergetic 140 MeV aluminum ion. Again, the stopping power calculations suggest more uniform heating of diamond foil with the laser-driven ion beam. In Fig. 3(b), a target depth of 21 µm is equivalent to the travel distance of ions within a 15 µm thick diamond at 45°.

In Figs. 3(a) and 3(b), the area under the curve defined by the data points in each plot represents the average energy lost within the target by a single incident aluminum ion, which is equivalent to the energy deposited to the target atoms. In the next section, we show how one can estimate the absorbed energy per atom using the stopping power calculations.

**III. Heating per atom and expected temperature calculations**

Figure 4 shows an example scenario that illustrates how we estimate the ion kinetic energy deposition into the foils. In this example, we assume ten aluminum ions are incident on a target that consists of ten million atoms. The average kinetic energy of the incident ions is 140 MeV, and each ion loses an average kinetic energy of 40 MeV while passing through the target. Then, one can calculate the average absorbed energy per target atom, or heating per atom, using the following relation [37]:

$$\text{Heating per atom} = \frac{N_{ion} <E_{deposit}>}{N_{target}}, \quad (1)$$

where $N_{ion}$ is the total number of incident aluminum ions on the target, $<E_{deposit}>$ is the average energy deposited by one aluminum ion, and $N_{target}$ is the total number of target atoms irradiated by the ion beam. Using Eq. (1), the heating is calculated to be 40 eV per atom in this example scenario. Radiation losses are insignificant and are neglected in Eq. (1) because the anticipated temperatures of the plasmas are of order several eV [37], well below the temperature where radiation losses are significant. Using Eq. (1) and the input data to SRIM shown in Fig. 2, we expect an average heating per atom of 38 eV for gold and 6.3 eV for diamond at a source-to-target distance of 2.37 mm used in Ref. [37].

We use SESAME EOS tables to estimate the expected plasma temperatures of gold and diamond from the heating per atom calculations. Figure 5 shows the expected plasma temperatures of gold and diamond as functions of heating per atom. The solid red line indicates the expected temperature of gold as a function of heating per atom using SESAME EOS table #2700, and dashed red line shows the expected temperature of gold when using #2705 EOS table [59]. At a heating of 38 eV/atom, there is a difference of 0.5 eV (~10%) in the temperatures predicted from two different EOS tables for gold. The difference between the expected temperatures when using different EOS tables was not as large for diamond. The solid blue line indicates the temperature of diamond using SESAME EOS table #7834, while the dashed blue line shows the expected temperature of diamond with SESAME #7830 table [46].

Figures 6(a) shows the expected temperature of gold as a function of target depth using SESAME EOS table #2700 (solid red circles with red error bars) and #2705 (hollow black circles with black error bars). Although both EOS tables predict uniform heating within the 10 μm thick gold foil (45° incidence angle has been taken into account in this figure), the resulting temperatures differ by 0.5 eV in Fig. 6(a). SESAME EOS table #2705 was produced in 2011 as a replacement for SESAME #2700 table [59], which was created in 1978. SESAME #2705 table gives predictions of the room-temperature isotherm, principal Hugoniot, thermal expansion, heat capacity, melt line, and vapor pressure for pure gold that are substantially different from and superior to the equivalent predictions using SESAME #2700 table [59].

Figure 6(b) shows similar calculations for a 15 μm thick diamond foil. Solid blue triangles indicate the expected temperature of diamond using SESAME #7834 table, while hollow red triangles with red error bars show the expected temperature using SESAME #7830 table. Both SESAME #7830 and #7834 tables predict nearly identical temperatures of diamond after heating.

The vertical error bars in Figs. 6(a) and 6(b) represent the uncertainties in the expected temperatures of gold and diamond, respectively, owing to the observed shot-to-shot variation of about ±30% in the incident number of aluminum ions [37]. To account for the variation in the number of incident aluminum ions, we have added ±30% error bars in the heating per atom calculations [37], which resulted in rather large error bars of up to ±1.0 eV in the expected temperatures of gold in Fig. 6(a) and error bars up to ±0.5 eV for diamond in Fig. 6(b). Since we estimate the errors in SRIM calculations to be smaller than 10% and heating from other sources such as laser-generated protons, x-rays, hot electrons to be insignificant, the error bars in Figs. 6(a) and 6(b) are mostly determined by the shot-to-shot variation in the ion fluence.

Based on calculations of electron-electron and electron-ion collision frequencies within our target [17,18], we expect that local thermal equilibrium is reached within several picoseconds, so the plasma temperature in Fig. 6(a) or in Fig. 6(b) represents both electron and ion temperatures. On the other hand, global thermal equilibrium is not expected to be reached within 10 μm thick gold or 15 μm thick diamond even after several hundreds of nanoseconds from heating based on our calculations of the diffusion coefficients of gold at 5.5 eV and diamond at 1.7 eV, which explains why the initial heating uniformity evidenced in Figs. 6(a) and 6(b) is important for the production of uniformly heated WDM sample.

In Figs. 7(a) and 7(b), we examine how sensitive the uniform heating conditions are to the energy spectrum of the incident aluminum ions by using different ion energy spectra in our SRIM simulations. Fig. 7(a) shows the calculated heating uniformity in the gold foil for 3 different input energy spectra of the aluminum ions. Solid red circles show the expected heating uniformity within the gold foil using the energy spectrum shown in Fig. 1 (black bars). Hollow blue triangles show the expected temperature of gold as a function of the target depth assuming each aluminum ion had 10% more kinetic energy, while hollow black squares show the expected temperature of gold assuming 10% less kinetic energy for each ion.

Figure 7(b) shows similar calculations of the expected temperature of diamond as a function of target depth using three different energy spectra. Solid blue triangles indicate the expected temperature of diamond as a function of target depth using the measured energy spectrum, while hollow red circles indicate the results using the ions with 10% more kinetic energy and hollow black squares indicate the results using the ions with 10% less kinetic energy.

In Figs. 7(a) and 7(b), we have adjusted the total number of incident aluminum ions so that the total kinetic energy of the incident ion beam is conserved. For the case with aluminum ions with 10% more kinetic energy, we assume the total number of incident ions is 9% less than the measured ion fluence. For the case with aluminum ions with 10% less kinetic energy, we assume 11% more ions are incident on the target. We have used SESAME EOS table #2705 for gold and SESAME #7834 table for diamond in converting the SRIM predictions of deposited energy to material temperatures in these simulations.

For applications involving materials other than gold and diamond, we have calculated the expected heating per atom for various target materials in Fig. 8(a). The same aluminum ion beam heats the target material, and error bars of ±15% are shown in the figure. We use ±15% errors instead of ±30% errors reported in Ref. [37] because we expect a better control of the ion beam in future experiments. The heating per atom increases almost monotonically with atomic number, primarily because of the increase in the total number of electrons in the target atom. We have also calculated expected plasma temperatures of various target materials using heating per atom calculations in Fig. 8(a) and their corresponding SESAME EOS tables. Figure 8(b) shows the expected plasma temperature as a function of the atomic number at a source-to-target distance of 2.37 mm. The two experimental data points in Ref. [37] are also shown in the figure. The plasma temperature generally increases with the atomic number, and there are multiple local minima around the location of the noble gases (Z=2, 10, 18, 36, 54, 86). We speculate that this is owing to the ionization potential peaks for the noble gases.

In Fig. 8 and throughout this paper, we have assumed the targets remain at solid density throughout the heating. This is a valid assumption because the volume changes during heating are expected to be small. We estimate the volume increase during heating to be at most 3% for gold and 2% for diamond based on the measured average expansion speeds in Ref. [37], and expect similar amount of volume increase during heating for other materials. The rise time of heating was calculated to be about 20 ps at a source-to-target distance of 2.37 mm in Ref. [37]. For a 10 μm thick gold foil, this translates to a volume increase of 3% during heating using the expansion speed of gold at 7.5 μm/ns, while a volume increase of 2% is expected for a 15 μm thick diamond expanding at 6.7 μm/ns.

## IV. Conclusions

We have presented a series of analyses that suggest uniform heating of gold and diamond with a laser-driven quasi-monoenergetic aluminum ion beam. Indeed, it appears that a quasi-monoenergetic ion beam with a small but finite energy spread can heat ~10 μm thick target foils more uniformly than a perfectly monoenergetic ion beam of the same mean energy. We have investigated the robustness of the heating uniformity by applying several different perturbations to the ion energy spectra used in our SRIM calculations. For future applications, we have presented the expected temperatures achievable by various target materials with the current experimental platform.

Such uniformly heated WDM samples would be useful for EOS, opacity, and conductivity [3,9] measurements. They can also be used for stopping power measurements of WDM [47]. Such targets would be also useful for validating our understanding of the physics of giant planet interiors, which are also in WDM state [43].

**Acknowledgments**

The authors would like to thank S. Crockett and K.G. Honnell for their advice regarding the use of SESAME tables, and thank B.M. Hegelich and G. Dyer at the University of Texas at Austin and L. Yin, S. Palaniyappan, and D.C. Gautier from LANL for valuable discussions. This work was performed at LANL, operated by Los Alamos National Security, LLC, for the U.S. DOE under Contract No. DE-AC52-06NA25396, and was supported in part by the LANL LDRD program.

**References**

[1]   W. S. Fann, R. Storz, H. W. K. Tom, and J. Bokor, Phys. Rev. Lett. **68**, 2834 (1992).
[2]   A. Forsman, A. Ng, G. Chiu, and R. M. More, Phys. Rev. E **58**, R1248 (1998).
[3]   K. Widmann, T. Ao, M. E. Foord, D. F. Price, A. D. Ellis, P. T. Springer, and A. Ng, Phys. Rev. Lett. **92**, 125002 (2004).
[4]   P. Patel *et al.*, Phys. Rev. Lett. **91**, 125004 (2003).
[5]   A. Mančić *et al.*, Phys. Rev. Lett. **104**, 035002 (2010).
[6]   A. Mancic *et al.*, High Energy Density Physics **6**, 21 (2010).
[7]   D. J. Hoarty, T. Guymer, S. F. James, E. Gumbrell, C. R. D. Brown, M. Hill, J. Morton, and H. Doyle, High Energy Density Physics **8**, 50 (2012).
[8]   M. Gauthier *et al.*, Phys. Rev. Lett. **110**, 135003 (2013).
[9]   Z. Chen, B. Holst, S. E. Kirkwood, V. Sametoglu, M. Reid, Y. Y. Tsui, V. Recoules, and A. Ng, Phys. Rev. Lett. **110**, 135001 (2013).
[10]  W. Bang *et al.*, Phys. Rev. Lett. **111**, 055002 (2013).
[11]  P. M. Leguay *et al.*, Phys. Rev. Lett. **111**, 245004 (2013).
[12]  M. Barbui *et al.*, Phys. Rev. Lett. **111**, 082502 (2013).

[13] U. Zastrau *et al.*, Phys. Rev. Lett. **112**, 105002 (2014).
[14] A. Lévy *et al.*, Phys. Plasmas **22**, 030703 (2015).
[15] K. U. Akli *et al.*, Phys. Rev. Lett. **100**, 165002 (2008).
[16] F. Perez *et al.*, Phys. Rev. Lett. **104**, 085001 (2010).
[17] S. M. Vinko *et al.*, Nature **482**, 59 (2012).
[18] C. Fourment, F. Deneuville, D. Descamps, F. Dorchies, S. Petit, O. Peyrusse, B. Holst, and V. Recoules, Phys. Rev. B **89**, 161110 (2014).
[19] W. Bang *et al.*, Phys. Rev. E **90**, 063109 (2014).
[20] W. Bang *et al.*, Phys. Rev. E **88**, 033108 (2013).
[21] W. Bang, G. Dyer, H. J. Quevedo, A. C. Bernstein, E. Gaul, M. Donovan, and T. Ditmire, Phys. Rev. E **87**, 023106 (2013).
[22] W. Bang, G. Dyer, H. J. Quevedo, A. C. Bernstein, E. Gaul, J. Rougk, F. Aymond, M. E. Donovan, and T. Ditmire, Phys. Plasmas **20**, 093104 (2013).
[23] W. Bang, Phys. Rev. E **92**, 013102 (2015).
[24] P. Norreys *et al.*, Nucl. Fusion **54**, 054004 (2014).
[25] A. Maksimchuk, S. Gu, K. Flippo, D. Umstadter, and V. Y. Bychenkov, Phys. Rev. Lett. **84**, 4108 (2000).
[26] R. Snavely *et al.*, Phys. Rev. Lett. **85**, 2945 (2000).
[27] S. C. Wilks *et al.*, Phys. Plasmas **8**, 542 (2001).
[28] M. Roth *et al.*, Phys. Rev. ST Accel. Beams **5**, 061301 (2002).
[29] B. M. Hegelich *et al.*, Nature **439**, 441 (2006).
[30] M. Schollmeier *et al.*, Phys. Rev. Lett. **101**, 055004 (2008).
[31] D. Jung *et al.*, Phys. Rev. Lett. **107**, 115002 (2011).
[32] T. Bartal *et al.*, Nat Phys **8**, 139 (2012).
[33] H. Daido, M. Nishiuchi, and A. S. Pirozhkov, Rep. Prog. Phys. **75**, 056401 (2012).
[34] B. M. Hegelich *et al.*, New J. Phys. **15**, 085015 (2013).
[35] D. Jung *et al.*, Phys. Plasmas **20**, 083103 (2013).
[36] A. Macchi, M. Borghesi, and M. Passoni, Rev. Mod. Phys. **85**, 751 (2013).
[37] W. Bang, B. J. Albright, P. A. Bradley, D. C. Gautier, S. Palaniyappan, E. L. Vold, M. A. S. Cordoba, C. E. Hamilton, and J. C. Fernández, Sci. Rep. **5**, 14318, doi:10.1038/srep14318 (2015).
[38] R. A. Snavely *et al.*, Phys. Plasmas **14**, 092703 (2007).
[39] E. Brambrink *et al.*, Phys. Rev. E **75**, 065401 (2007).
[40] G. Dyer *et al.*, Phys. Rev. Lett. **101**, 015002 (2008).
[41] A. Pelka *et al.*, Phys. Rev. Lett. **105**, 265701 (2010).
[42] M. Koenig *et al.*, Plasma Phys. Controlled Fusion **47**, B441 (2005).
[43] R. F. Smith *et al.*, Nature **511**, 330 (2014).
[44] C. R. D. Brown *et al.*, Sci. Rep. **4**, 5214 (2014).
[45] T. G. White *et al.*, Phys. Rev. Lett. **112**, 145005 (2014).
[46] K. Falk, E. J. Gamboa, G. Kagan, D. S. Montgomery, B. Srinivasan, P. Tzeferacos, and J. F. Benage, Phys. Rev. Lett. **112**, 155003 (2014).
[47] A. B. Zylstra *et al.*, Phys. Rev. Lett. **114**, 215002 (2015).
[48] D. Li, H. Liu, S. Zeng, C. Wang, Z. Wu, P. Zhang, and J. Yan, Sci. Rep. **4**, 5898 (2014).
[49] Y. Sentoku, I. Paraschiv, R. Royle, R. C. Mancini, and T. Johzaki, Phys. Rev. E **90**, 051102 (2014).
[50] H. Schwoerer, S. Pfotenhauer, O. Jackel, K. U. Amthor, B. Liesfeld, W. Ziegler, R. Sauerbrey, K. W. D. Ledingham, and T. Esirkepov, Nature **439**, 445 (2006).
[51] C. A. J. Palmer *et al.*, Phys. Rev. Lett. **106**, 014801 (2011).
[52] S. Kar *et al.*, Phys. Rev. Lett. **109**, 185006 (2012).

[53] D. Haberberger, S. Tochitsky, F. Fiuza, C. Gong, R. A. Fonseca, L. O. Silva, W. B. Mori, and C. Joshi, Nat Phys **8**, 95 (2012).
[54] S. Palaniyappan, C. Huang, D. C. Gautier, C. E. Hamilton, M. A. Santiago, C. Kreuzer, R. C. Shah, and J. C. Fernandez, arXiv:1506.07548 (2015).
[55] J. F. Ziegler, M. D. Ziegler, and J. P. Biersack, Nucl. Instrum. Methods Phys. Res. B **268**, 1818 (2010).
[56] H. Paul and D. Sánchez-Parcerisa, Nucl. Instrum. Methods Phys. Res. B **312**, 110 (2013).
[57] S. P. Lyon and J. D. Johnson, Los Alamos National Laboratory Report No. LA-UR-92-3407 (1992).
[58] D. Jung *et al.*, Rev. Sci. Instrum. **82**, 043301 (2011).
[59] J. Boettger, K. G. Honnell, J. H. Peterson, C. Greeff, and S. Crockett, AIP Conf. Proc. **1426**, 812 (2012).

**Figures**

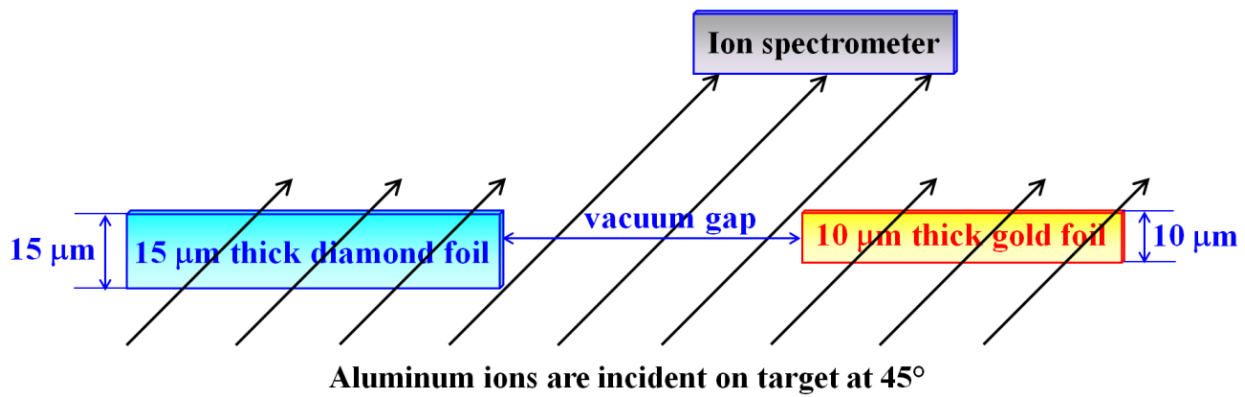

FIG. 1. (Color online) Schematic layout of the target (not to scale). A laser-driven beam of quasi-monoenergetic aluminum ions is incident on the target at 45°. The diamond and gold foils are heated isochorically by the energetic aluminum ions, and become WDM. The ions that pass through the vacuum gap between the two foils are recorded by a magnetic ion spectrometer. On some shots, a Thomson parabola ion spectrometer replaced the magnetic ion spectrometer.

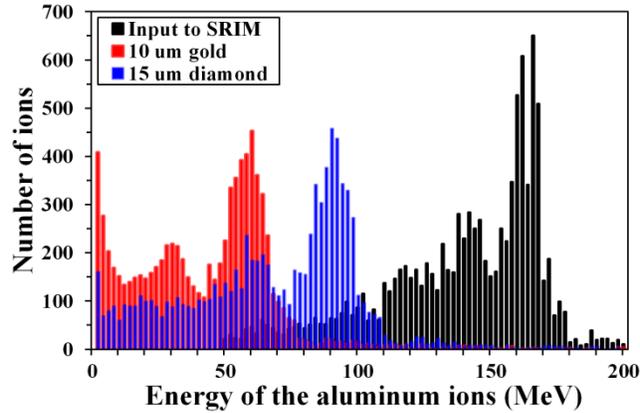

FIG. 2. (Color online) Energy spectrum of the incident aluminum ions at 0° (black bars) measured from Ref. [54] along with the calculated ion energy spectra after penetrating through a 10 μm gold foil (red bars) or a 15 μm diamond foil (blue bars). The measured ion energy spectrum (indicated as black bars) was used as input data in our SRIM calculations.

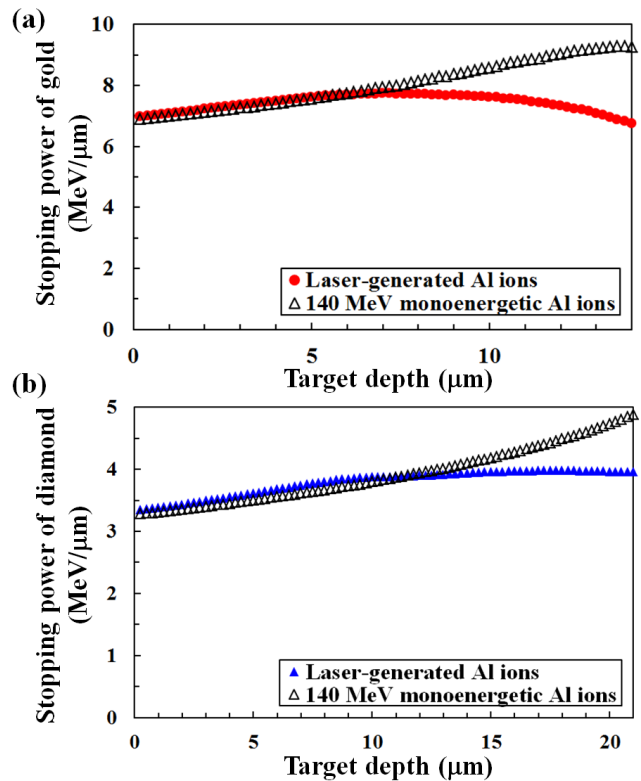

FIG. 3. (Color online) (a) The average stopping power of gold for laser-generated aluminum ions (solid red circles) is shown as a function of target depth. The stopping power of gold for a perfectly monoenergetic 140 MeV aluminum ion is shown together (hollow black triangles) to emphasize the heating uniformity expected from the laser-driven ion beam. (b) Similar calculations of the average stopping power of diamond for laser-generated aluminum ions (solid blue triangles) and for a 140 MeV aluminum ion (hollow black triangles) are shown as functions of the target depth.

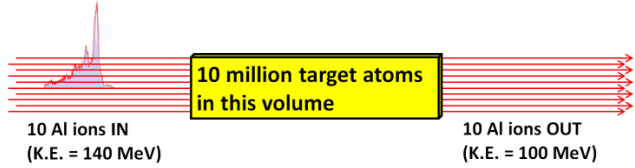

FIG. 4. (Color online) An example scenario illustrates how we calculate the heating per atom. Ten aluminum ions with an average kinetic energy of 140 MeV are incident on a target. In this example, the target consists of ten million atoms, and the aluminum ions exit the target with an average kinetic energy loss of 40 MeV per ion.

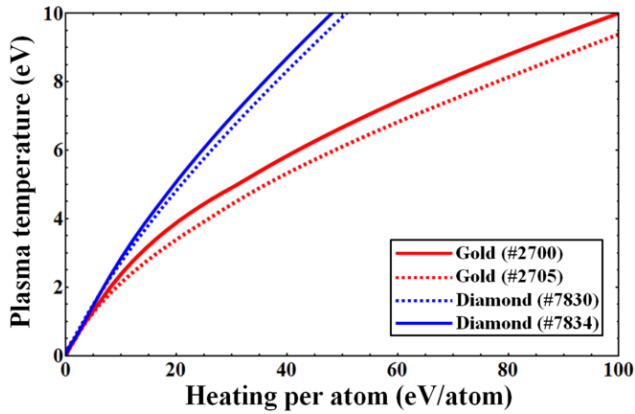

FIG. 5. (Color online) Expected plasma temperatures of gold and diamond are shown as functions of the heating per atom. SESAME EOS #2700 (solid red line) and #2705 (dashed red line) are used for gold, and SESAME EOS #7830 (dashed blue line) and #7834 (solid blue line) are used for diamond.

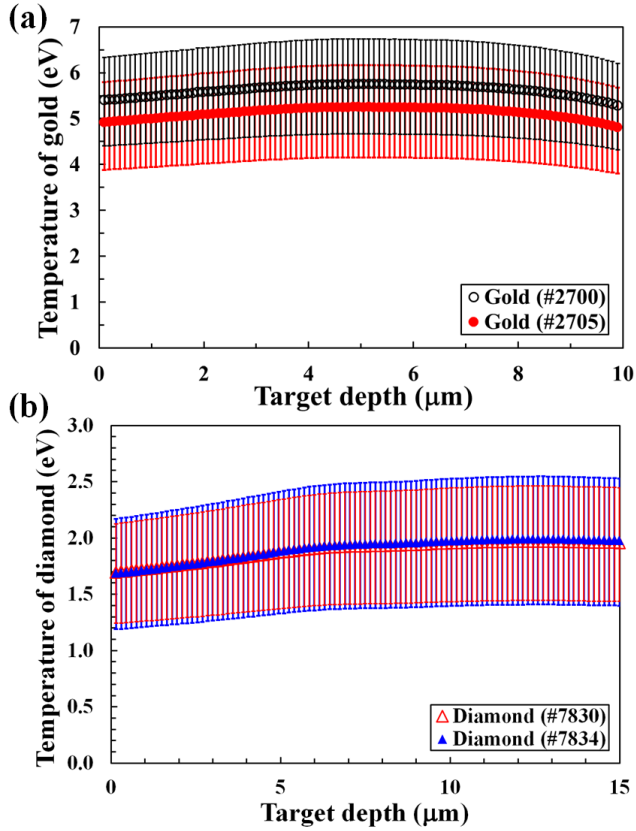

FIG. 6. (Color online) (a) Expected plasma temperature of a 10 μm thick gold foil heated by the quasi-monoenergetic aluminum ion beam is shown as a function of the target depth. SESAME EOS #2700 (hollow black circles) and #2705 (solid red circles) were used. (b) Expected plasma temperature of a 15 μm thick diamond foil is shown as a function of the target depth. SESAME EOS #7830 (hollow red triangles) and #7834 (solid blue triangles) were used. The error bars were calculated assuming a shot-to-shot variation of ±30% in the incident aluminum ion fluence.

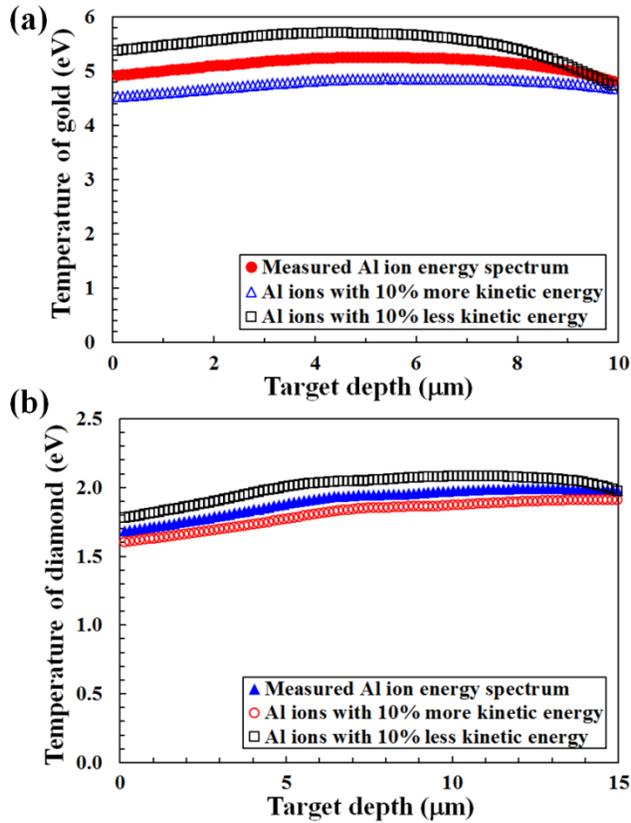

FIG. 7. (Color online) (a) Calculated heating uniformity in the gold foil for 3 different input energy spectra of the aluminum ions. Solid red circles show the expected heating uniformity within the gold foil using the measured energy spectrum of the incident aluminum ions in SRIM. Hollow blue triangles show the resulting temperatures of gold versus target depth assuming each aluminum ion had 10% more kinetic energy than the measured energy, while hollow black squares show the temperatures assuming 10% less kinetic energy for each ion. (b) Calculated temperatures of diamond are shown as functions of the target depth using 3 different energy spectra. Solid blue triangles indicate the temperatures obtained from the measured spectrum, while hollow red circles indicate the results using the ions with 10% more kinetic energy and hollow black squares indicate the results using the ions with 10% less kinetic energy. SESAME table #2705 is used for gold and #7834 table is used for diamond.

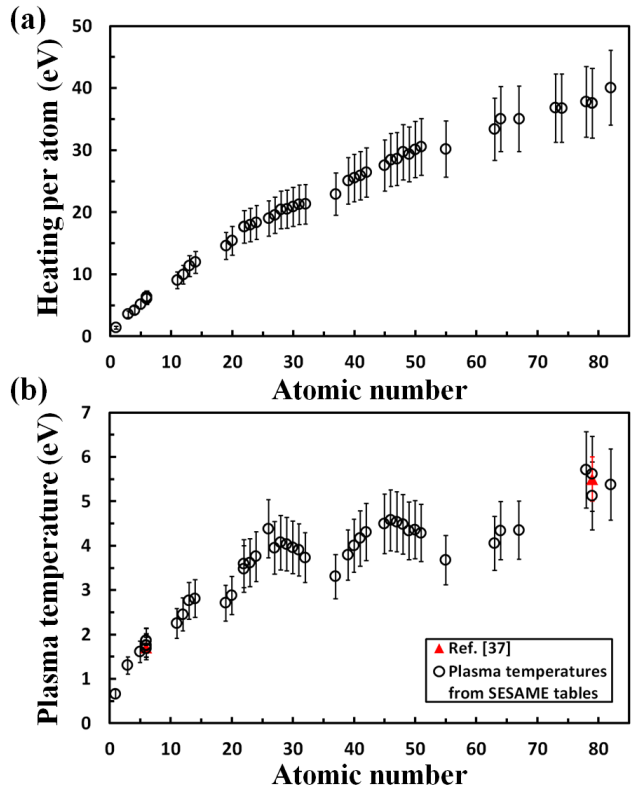

FIG. 8. (Color online) (a) Heating per atom as a function of the atomic number. The expected heating per atom calculations with the quasi-monoenergetic aluminum ion beams are shown for various target materials to serve as a reference for future experiments. The heating per atom increases with the atomic number. (b) Expected plasma temperatures of various target materials are shown as a function of the atomic number. The temperatures determined by the measured expansion speeds in Ref. [37] are shown together with the calculated plasma temperatures of various target materials using the heating per atom calculations and their corresponding SESAME EOS tables. Error bars of ±15% are shown in the above figures.